\documentclass[twocolumn,english,revtex]{revtex4-1}
\usepackage[T1]{fontenc}
\usepackage[latin9]{inputenc}
\usepackage{amsmath}
\usepackage{amssymb}
\usepackage{graphicx}
\usepackage{esint}
\usepackage{babel}
\begin{document}

\title{\noindent Squeezed thermal reservoirs as a resource for a nano-mechanical
engine\\
 beyond the Carnot limit}

\author{\noindent Jan Klaers$^{1}$}

\email{jklaers@phys.ethz.ch}

\selectlanguage{english}%

\author{\noindent Stefan Faelt$^{1}$, Atac Imamoglu$^{1}$, and Emre Togan$^{1}$}

\address{\noindent $^{1}$Institute for Quantum Electronics, ETH Z\"urich,
CH-8093 Z\"urich, Switzerland}
\begin{abstract}
\noindent The efficient conversion of thermal energy to mechanical
work by a heat engine is an ongoing technological challenge. Since
the pioneering work of Carnot, it is known that the efficiency of
heat engines is bounded by a fundamental upper limit \nobreakdash-
the Carnot limit. Theoretical studies suggest that heat engines may
be operated beyond the Carnot limit by exploiting stationary, non-equilibrium
reservoirs that are characterized by a temperature as well as further
parameters. In a proof-of-principle experiment, we demonstrate that
the efficiency of a nano-beam heat engine coupled to squeezed thermal
noise is not bounded by the standard Carnot limit. Remarkably, we
also show that it is possible to design a cyclic process that allows
for extraction of mechanical work from a single squeezed thermal reservoir.
Our results demonstrate a qualitatively new regime of non-equilibrium
thermodynamics at small scales and provide a new perspective on the
design of efficient, highly miniaturized engines.
\end{abstract}
\maketitle

\section{\hspace{-2mm}Introduction}

\vspace{-2mm}

Advances in micro- and nano-technology allow for testing concepts
derived from classical thermodynamics in regimes where the underlying
assumptions, such as the thermodynamic limit and thermal equilibrium,
no longer hold \cite{Liphardt2002,Toyabe2010,Berut2012,Trotzky2012,Roldan2014,Vinjanampathy2015}.
Extremely miniaturized forms of heat engines, where the working medium
is represented by a single particle, have revealed a fluctuation-dominated
regime in the conversion of heat to work far away from the thermodynamic
limit \cite{Blickle2012,Koski2014,Martinez2016,Rossnagel2016,Krishnamurthy2016}.
By employing non-equilibrium reservoirs it is furthermore expected
that the efficiency of work generation surpasses the standard Carnot
limit \cite{Carnot1978}, as has been theoretically suggested for
quantum coherent \cite{Scully2003}, quantum correlated \cite{Dillenschneider2009,Perarnau-Llobet2015}
and squeezed thermal reservoirs \cite{Huang2012,Rossnagel2014,Correa2014,Niedenzu2016,Niedenzu2017}. 

In optics, the electric field of a monochromatic wave can be decomposed
into two quadrature components that vary as $\cos\omega t$ and $\sin\omega t$
respectively. For coherent states such as laser light, the uncertainties
in the two quadratures are equal and follow the lower bound of Heisenberg's
uncertainty relation. By contrast, squeezed states of light have reduced
fluctuations in one (squeezed) quadrature at the cost of enhanced
fluctuations in the other (anti-squeezed) quadrature allowing for
optical measurements with reduced quantum noise \cite{LIGO2011,Wolfgramm2010}.
Squeezed states are, however, neither restricted to electromagnetic
waves nor to minimum uncertainty states. For example, a mechanical
oscillator may be prepared in a squeezed thermal state \cite{Fearn1988,Kim1989,Tucci1991}
by a periodic modulation of the spring constant \cite{Rugar1991}.
This results in a state with reduced thermal fluctuations in one quadrature
(e.g. momentum) and enhanced thermal fluctuations in the other quadrature
(e.g. position) compared to the expected level of fluctuations at
the given temperature. In the context of heat engines, a theoretical
work by Roßnagel et al. \cite{Rossnagel2014} suggests that squeezed
thermal states may be used as an additional resource to overcome the
standard Carnot limit. Due to the non-equilibrium nature of squeezed
thermal reservoirs, this result does not violate the second law of
thermodynamics. In our work, we present a physical realization of
such an engine with a working medium consisting of a vibrating nano-beam
that is driven by squeezed electronic noise to perform work beyond
the standard Carnot limit. Furthermore, we demonstrate that by a phase-selective
coupling to the squeezed or anti-squeezed quadrature, work can be
extracted even from a single squeezed reservoir, which is not possible
with a standard thermal reservoir \cite{Scully2003,Abah2015-1}. 

\vspace{-3mm}

\section{\hspace{-2mm}Nano-beam heat engine}

\vspace{-3mm}

Figure 
\begin{figure}
\noindent \begin{centering}
\includegraphics[width=0.47\textwidth]{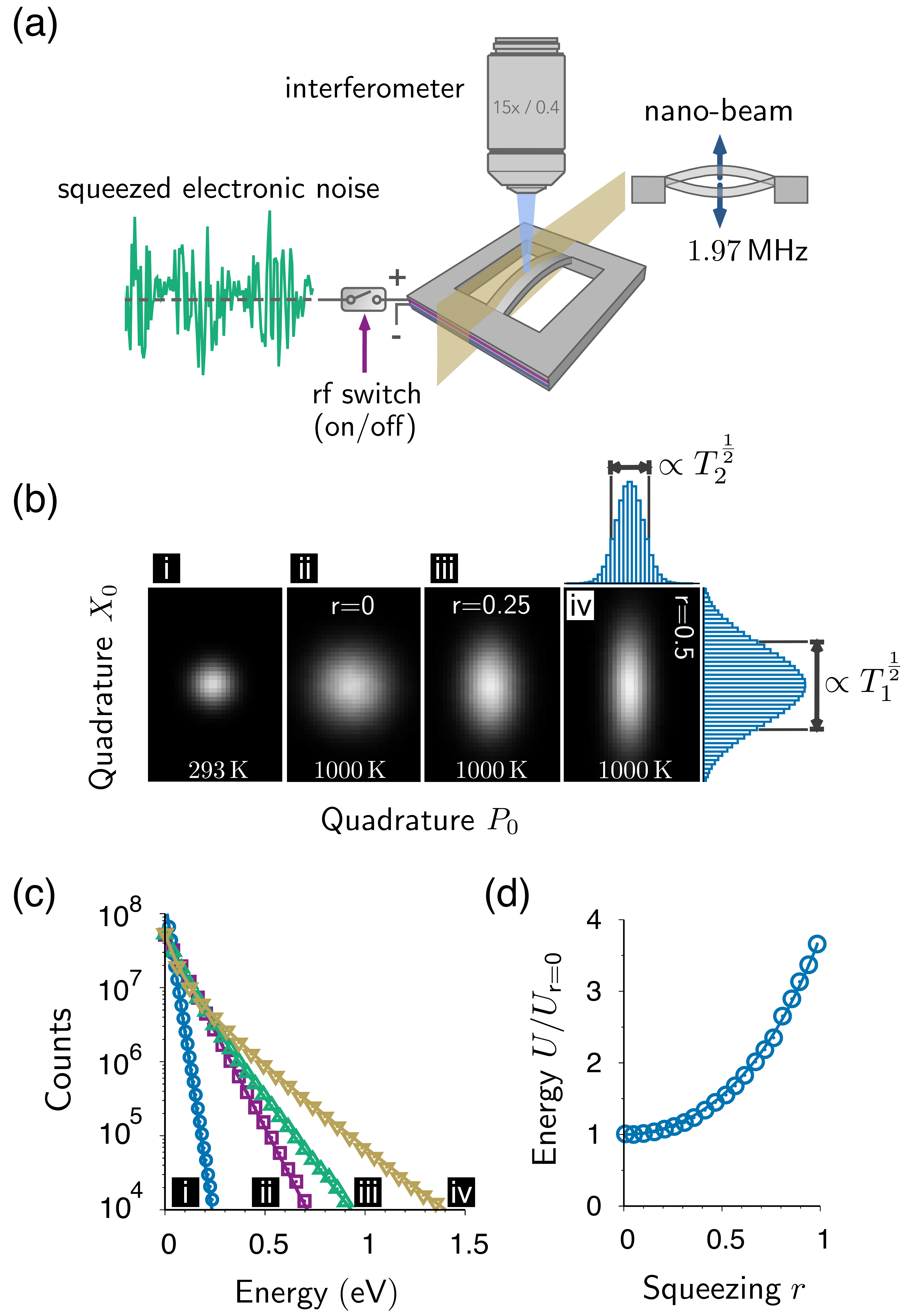}
\par\end{centering}

\caption{Nano-beam heat engine. (a) Doubly-clamped nano-beam piezo-electrically
coupled to squeezed electronic noise. (b) Phase space density of the
nano-beam motion (in rotating frame), (i) when no additional noise
is applied, (ii)-(iv) when (squeezed) noise is applied. The fluctuations
in position and momentum are characterized by two temperature parameters
$T_{1,2}$ proportional to the Gaussian variance of the line integrated
probability densities. The temperature is $T=293\,\text{K}$ for (i),
and $T=\sqrt{T_{1}T_{2}}\simeq1000\,\text{K}$ for (ii-iv). The squeezing
parameter follows $r=0$ for (i-ii), and $r=\ln(T_{1}/T_{2})/4=0.25,\,0.5$
for (iii-iv). (c) Measured energy histogram (symbols) of the nano-beam
motion in a (squeezed) thermal state. The solid lines show the theoretically
expected distributions. Experimental parameters as in Fig. 1b. (d)
Average energy of the nano-beam motion (circles) as a function of
the squeezing parameter $r$. The energies are normalized to the energy
at vanishing squeezing $U_{r=0}=k_{\text{B}}T$, with the temperature
of the system being kept fixed. The solid line corresponds to the
theoretical expectation $U/U_{r=0}=1+2\sinh^{2}r$. Statistical errors
(s.e.m.) are smaller than the symbol size for all data points.}
\end{figure}
 1a shows a sketch of the experiments: The working medium of our heat
engine consists of a single harmonic oscillator given by the fundamental
flexural mode of a doubly-clamped (GaAs) nano-beam structure with
eigenfrequency $\nu=\omega/2\pi=1.97\,\text{MHz}$, quality factor
of order $Q\simeq10^{3}$ at room temperature and under vacuum conditions
($p\simeq10^{-4}\,\text{mbar}$). The beam has a length of $18.8\,\mu\text{m}$,
width of $2\,\mu\text{m}$, thickness of $270\,\text{nm}$, and is
fabricated using conventional nano-structuring techniques such as
electron beam lithography and selective etching. In the growth direction
the beam contains two doped layers, which allow us to apply electric
fields across the beam material. These electric fields lead to forces
being applied to the beam structure due to the piezo-electricity of
the employed gallium arsenide material. When driven by a noisy waveform,
this creates a random force that allows us to mimic an engineered
thermal environment for the beam structure. The waveform is synthesized
from two independent white noise signals $\xi_{1,2}(t)$ that are
mixed with sine and cosine component of a phase reference at frequency
$\nu$ leading to a stochastic force $f(t)=a_{0}\left[\text{e}^{+\tilde{r}}\xi_{1}(t)\,\cos(\omega t)+\text{e}^{-\tilde{r}}\xi_{2}(t)\,\sin(\omega t)\right]$.
A positive squeezing parameter $\tilde{r}$ corresponds to an amplified
cosine and attenuated sine component at frequency $\nu$ in the thermal
bath, while the overall strength of the noise can be controlled by
the amplitude $a_{0}$.

To demonstrate the generation of squeezed thermal states, we experimentally
determine the motional state of the nano-beam by recording the instantaneous
position (out-of-plane displacement) of the beam via Mach-Zehnder
interferometry \cite{Klaers2016} and calculating the corresponding
phase-space probability distribution, see Fig. 1b. In the absence
of additional noise ($a_{0}=0$), the nano-beam motion is fully determined
by the residual thermal noise at room temperature, see panel (i) in
Fig. 1b. By increasing the noise amplitude $a_{0}$, the state of
the nano-beam can be prepared in a thermal state at higher temperature
$T=1000\,\text{K}$, see panel (ii). Squeezed noise leads to the expected
elliptical phase-space probability distribution, see panels (iii)
and (iv). We emphasize that all results are presented in a position-momentum-frame
that rotates with frequency $\nu$ with respect to the laboratory
frame. The observed probability densities closely follow the theoretically
expected Gaussian distribution \cite{Fearn1988}

\noindent 
\begin{equation}
\rho(x_{0},p_{0})\,\propto\,\exp\left(-\frac{\hbar\omega x_{0}^{2}}{2k_{\text{B}}T_{1}}-\frac{\hbar\omega p_{0}^{2}}{2k_{\text{B}}T_{2}}\right)\;\text{,}
\end{equation}

\noindent where $x_{0}$, $p_{0}$ are dimensionless position and
momentum variables and $T_{1}$, $T_{2}$ are two temperature parameters
describing the level of fluctuations in the anti-squeezed and squeezed
quadratures (proportional to the variance of the gaussian distribution).
These parameters are related to the effective temperature and squeezing
parameter of the system by $T_{1,2}=T\,\exp(\pm2r)$ \cite{Tucci1991,Fearn1988}.
The corresponding energy histograms reveal exponential distributions
for cases (i) (circles) and (ii) (boxes) as is expected from purely
thermal states (Fig. 1c). The non-exponential decays for parameter
sets (iii) (upright triangles) and (iv) (upside-down triangles) demonstrate
the non-equilibrium nature of the state of the nano-beam. The solid
lines in Fig. 1c show the theoretically expected energy distribution
$\rho(E)\propto I_{0}\left(E\,\sinh(2r)\,/k_{\text{B}}T\right)\,\exp\left(-E\,\cosh(2r)\,/k_{\text{B}}T\right)$
with $I_{0}$ as the modified Bessel function of order zero \cite{Klaers2016}.
Finally, the mean energy $U$ of the nano-beam motion as a function
of the squeezing parameter $r$ (at fixed temperature) closely follows
the caloric equation of state $U=k_{\text{B}}T\,(1+2\sinh^{2}r)$,
as demonstrated in Fig. 1d.

\noindent 
\begin{figure*}
\noindent \begin{centering}
\includegraphics[width=1\textwidth]{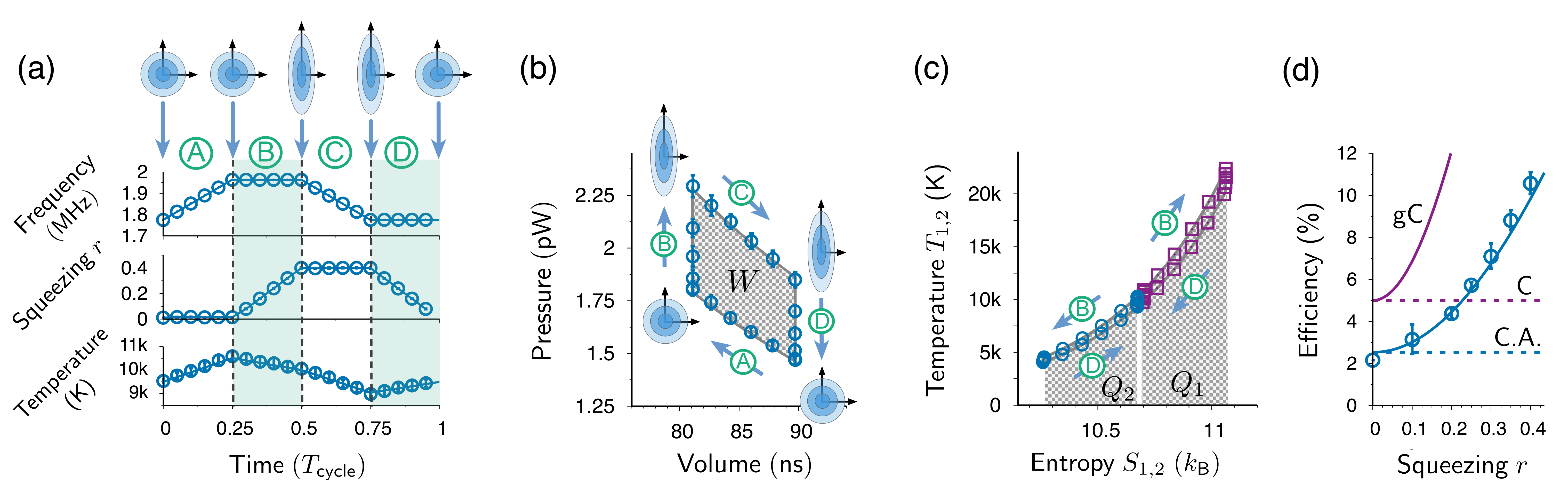}
\par\end{centering}

\caption{Otto cycle between a cold thermal and a hot squeezed thermal reservoir.
(a) Frequency, squeezing and temperature of the nano-beam motion throughout
one cycle of the engine ($T_{\text{cycle}}$ amounts to several seconds).
The shown protocol implements an Otto cycle at maximum power between
a cold thermal reservoir at $T_{\text{c}}=9,500\,\text{K}$ and a
hot squeezed reservoir at $T_{\text{h}}=10,000\,\text{K}$ with squeezing
parameter $r=0.4$. The four consecutive strokes include an isentropic
compression (A), isochoric heat addition (B), isentropic expansion
(C) and isochoric heat rejection (D). (b) Pressure-volume diagram
of the Otto cycle. The shaded area corresponds to the total work output
$W$ performed by the engine. The solid line represents the theoretically
expected behavior. (c) Temperature-entropy diagram of the Otto cycle.
$T_{1}$ ($T_{2}$) and $S_{1}$ ($S_{2}$) denote temperature and
entropy of the anti-squeezed (squeezed) quadrature. The solid line
shows the theoretical expectations. (d) Efficiency of the nano-beam
heat engine as a function of the squeezing parameter of the hot reservoir
(circles). The solid blue line shows the theoretical expectation $\eta=1-\sqrt{T_{\text{c}}/T_{\text{h}}}\,/\cosh(r)$.
The measured engine efficiency surpasses the standard Carnot (C) and
Curzon-Ahlborn (C.A.) efficiency for finite squeezing parameters (dashed
horizontal lines), but obeys a generalized Carnot limit (gC) \cite{Abah2015-1,Rossnagel2014}.
All error bars indicate statistical errors (s.e.m.) except for Fig.
2d (see Ref. \cite{Klaers2016} for details).}
\end{figure*}

In our experiment, the eigenfrequency of the fundamental flexural
mode can be tuned over a few $100\,\text{kHz}$ by applying a DC electrical
potential to the nano-beam \cite{Klaers2016} allowing for a cyclic
process with work output. The extraction of work in an engine is normally
realized by an increase in volume of a gaseous working medium driving
a piston. For a harmonically trapped working medium, such as the one
investigated here, an expansion of the working medium may be realized
by a decrease in trapping frequency $\omega$. The latter suggests
to define the volume of the system as the inverse trapping frequency
$V=\omega^{-1}$, see Ref. \cite{Romero-Rochin2005}. With this, a
relation similar to the state functional formulation of the first
law of thermodynamics holds \cite{Klaers2016} 

\noindent 
\begin{equation}
dU=T_{1}dS_{1}+T_{2}dS_{2}-pdV\;\text{,}
\end{equation}

\noindent with $S_{1}=-k_{\text{B}}\int_{-\infty}^{+\infty}\rho(x_{0})\ln(\rho(x_{0}))\,dx_{0}$
as the entropy of the anti-squeezed quadrature and the corresponding
definition for $S_{2}$ being obtained by replacing $x_{0}$ with
$p_{0}$. The pressure is defined by $p=-(\partial U/\partial V)_{S_{1},S_{2}}$,
which evaluates to $p=U/V$ for the given system. In this way, the
term $-p\,dV$ represents the work associated to a change in volume
or trapping frequency, whereas the terms $T_{i}\,dS_{i}$ with $i=1,2$
describe heat exchange with the environment. 

\vspace{0mm}

\section{\hspace{-2mm}Otto cycle with squeezed \ thermal reservoirs\ }

\vspace{-3mm}

\begin{figure*}
\noindent \begin{centering}
\includegraphics[width=0.95\textwidth]{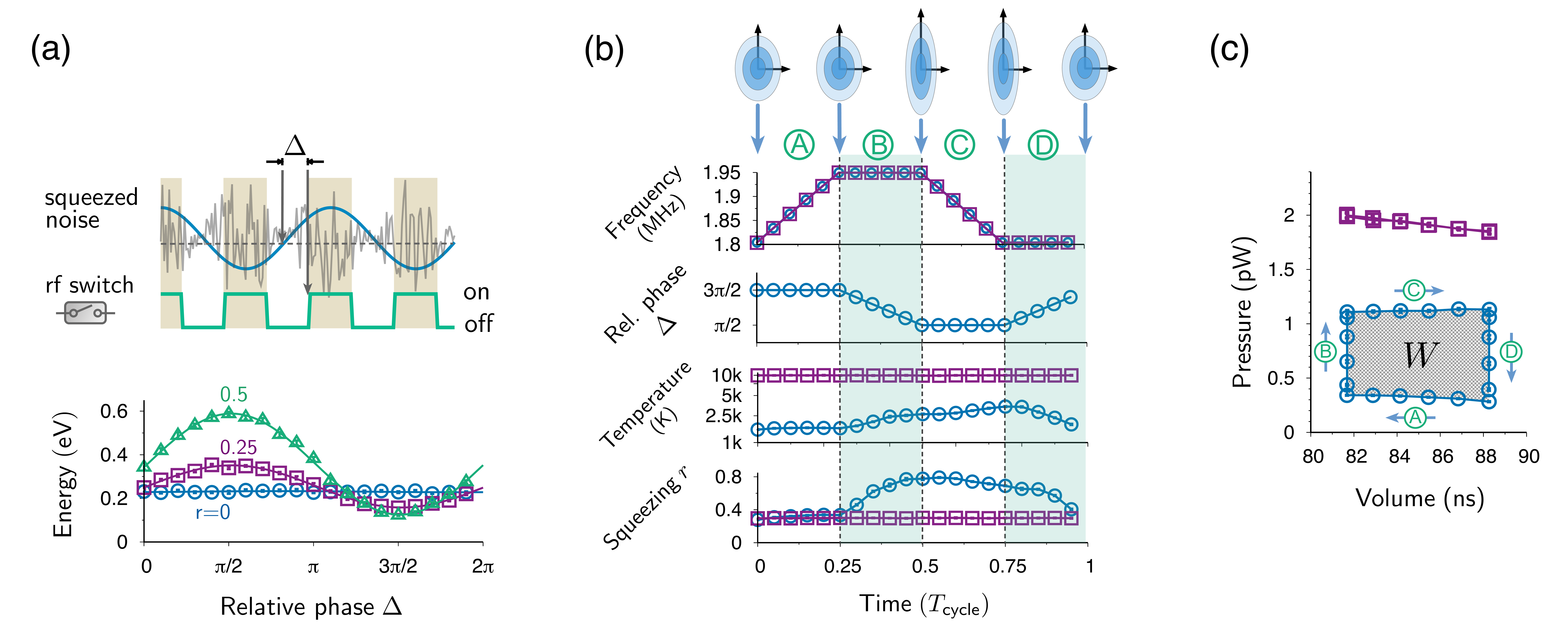}
\par\end{centering}

\caption{Work extraction from a single squeezed thermal reservoir. (a) By means
of a radio-frequency switch (rf switch in Fig. 1a) the coupling between
the squeezed electronic noise and the nano-beam is periodically switched
on and off with the phase difference between the phase of the squeezed
noise and the switching being denoted by $\Delta$ (top). Mean energy
of the working medium coupled to a squeezed bath as a function of
the relative phase $\Delta$ for three squeezing parameters $r=0,\,0.25,\,0.5$
(bottom). The data points are interpolated by harmonic functions with
varying amplitude and offset (solid lines). (b) Protocol to extract
work from a single squeezed reservoir with $T=10,000\,\text{K}$ and
$r=0.3$ by periodically varying mechanical eigenfrequency $\nu$
and phase $\Delta$ (circles in upper two graphs). The four strokes
include an isophasal compression, during which the phase relation
between squeezed noise and periodic coupling is kept fixed, an isochoric
heat addition, isophasal expansion, and isochoric heat rejection.
Temperature and squeezing of the nano-beam motional state as consequence
of the applied protocol (circles in lower two graphs). The case of
an unmodulated coupling is shown with box symbols. Lines between data
points are guides to the eye. (c) Pressure-volume diagram revealing
a finite amount of work ($W\simeq37\,\text{meV}$) extracted from
the single squeezed reservoir per cycle (circles). No work output
is observed for unmodulated coupling (boxes). Lines between data points
are guides to the eye. Statistical errors (s.e.m.) are smaller than
the symbol size for all data points.}
\end{figure*}

In a first line of experiments, we construct an Otto cycle between
a hot squeezed thermal reservoir at $T_{\text{h}}=10,000\,\text{K}$
with squeezing factor $r=0.4$ and a cold purely thermal bath at $T_{\text{c}}=9,500\,\text{K}$
under maximum power condition \cite{Rossnagel2014}. The four strokes
include an adiabatic compression, isochoric heat addition, adiabatic
expansion and isochoric heat rejection. Similar to colloidal heat
engines \cite{Martinez2016}, the working medium in our system cannot
be fully decoupled from its environment rendering the implementation
of adiabatic steps not obvious. Experimentally, it is however feasible
to replace the adiabatic steps by isentropic steps. The isentropic
steps are implemented by simultaneously varying frequency and temperature
such that the ratios $T_{1,2}/\omega$ remain constant, which conserves
the entropies $S_{1,2}$ \cite{Klaers2016}. Figure 2a shows frequency,
squeezing and temperature of the nano-beam state throughout one cycle
of the engine under quasi-static operation (the cycle time $T_{\text{cycle}}$
amounts to several seconds). The pressure-volume (p-V) and temperature-entropy
(T-S) diagrams are shown in Fig. 2b,c. The shaded area in the p-V
diagram corresponds to a net work output of $W\simeq26\,\text{meV}$
performed per cycle. We point out that presently, this work is neither
used to perform a certain task, nor stored in some form of potential
energy. The T-S diagram is composed of two closed curves corresponding
to temperature and entropy variation of the squeezed (circles) and
anti-squeezed quadrature (boxes) within one cycle. Note that the curve
for the anti-squeezed quadrature is run through clockwise, while the
curve for the squeezed quadrature is run through counterclockwise.
The net amount of heat consumed from the environment $Q_{\text{h}}$
during the hot isochore (B) corresponds to the difference of the two
shaded areas, which is $Q_{\text{h}}=Q_{1}-Q_{2}\simeq244\,\text{meV}$.
The efficiency of the cycle finally evaluates to $\eta=W/Q_{\text{h}}=(10.6\pm0.5)\text{\%}$,
which is roughly twice the efficiency of a Carnot cycle operating
between $T_{\text{h}}$ and $T_{\text{c}}$ and four times the Curzon-Ahlborn
efficiency \cite{Curzon1975}. We emphasize that the costs of providing
a squeezed thermal bath are not accounted for in our analysis, which
is fully analogous to neglecting the costs of providing a hot or cold
bath in standard thermodynamics. Figure 2d shows the efficiency for
varying squeezing parameters of the hot bath. The solid blue line
corresponds to the theoretical expectation $\eta=1-\sqrt{T_{\text{c}}/T_{\text{h}}}\,/\cosh(r)$,
see Ref. \cite{Rossnagel2014,Klaers2016}, whereas the two horizontal
dashed lines show the (standard) Carnot (C) and Curzon-Ahlborn (C.A.)
efficiencies given by $\eta_{\text{Carnot}}=1-T_{\text{c}}/T_{\text{h}}$
and $\eta_{\text{C.A.}}=1-\sqrt{T_{\text{c}}/T_{\text{h}}}$ respectively.
These results clearly demonstrate that squeezing, given as a free
resource, can be used to increase the work output and efficiency of
an engine beyond the (standard) Carnot limit. The engine operation,
however, does obey a generalized Carnot limit (solid line 'gC') \cite{Abah2015-1,Rossnagel2014}.

\vspace{-3mm}

\section{\hspace{-3mm}Work extraction from a single reservoir}

\vspace{-3mm}

In a second line of experiments, we eliminate the cold thermal heat
bath in the operation of the engine and introduce a phase-selective
thermal coupling that will allow to periodically extract work from
a single squeezed heat bath. In these measurements, the coupling of
the mechanical oscillator to the squeezed electronic noise is periodically
switched on and off by means of a rectangular control signal with
frequency $2\nu$ and duty cycle 50\%, see top panel in Fig. 3a, that
is applied to a radio-frequency switch. The mean energy of the mechanical
oscillator as a function of the relative phase $\Delta$ between the
phase of the switching function and the phase of the squeezed bath
is shown in the bottom panel of Fig. 3a for three different squeezing
parameters. For non-vanishing squeezing parameters (boxes and triangles),
the measured energy reaches a maximum when the oscillator is coupled
to the anti-squeezed quadrature of the bath at $\Delta_{\text{max}}\simeq\pi/2$,
whereas an energy minimum is obtained for coupling to the squeezed
quadrature at $\Delta_{\text{min}}\simeq3\pi/2$. This phase-selective
coupling allows us to extract work from a single squeezed reservoir
by a periodic variation of mechanical frequency and phase difference
$\Delta$ as shown in the upper two panels of Fig. 3b (circles). The
four strokes include an ``isophasal'' compression, during which
the phase relation between squeezed noise and periodic coupling is
kept fixed, an isochoric heat addition, isophasal expansion, and isochoric
heat rejection. Temperature and squeezing of the nano-beam state as
consequence of the applied protocol are shown in the bottom two panels
of Fig. 3b (circles). Finally, the p-V diagram in Fig. 3c clearly
reveals a finite amount of work ($W\simeq37\,\text{meV}$) that is
extracted from the single squeezed reservoir per cycle (circles).
Compared to the case of an unmodulated coupling (switch in 'on' position),
resulting in a squeezed thermal state with $T=10,000\,\text{K}$ and
$r=0.3$ throughout the cycle (boxes in Fig. 3b, c), the working medium
operates at a reduced temperature. The latter originates from the
incomplete decoupling of the oscillator from its environment during
the 'off'-phases of the phase-selective coupling. Here, an improved
isolation would further increase the extracted work per cycle.

\vspace{-5mm}

\section{\hspace{-2mm}Conclusions}

\vspace{-3mm}

Single-particle heat machines provide an excellent platform to test
theoretical advances in and our understanding of thermodynamics at
the micro- and nano-scale. In this work, a minimalist heat engine
has been realized that takes advantage of squeezed heat to outperform
conventional heat engines. The non-equilibrium nature of these reservoirs
permit work extraction from a single reservoir and engine efficiencies
unbounded by the standard Carnot limit. A major open question remains
whether highly miniaturized heat engines, such as the one reported
here, will be able to use the extracted work to accomplish mesoscopic
tasks as transporting particles \cite{H=0000E4nggi2009} or manipulating
biological matter \cite{Douglas2012}. Squeezed thermal noise may
naturally arise in non-equilibrium environments with temporally or
spatially modulated temperature profiles and could be used to fuel
such devices.

\end{document}